\documentclass[preprint,12pt,number]{elsarticle}

\usepackage{amssymb}
\usepackage{amsmath}
\usepackage{subfig}

\journal{Nuclear Physics B}

\begin{document}

\begin{frontmatter}



\title{Search for long-lived charged particles using the CMS detector in Run-2}

\author{Tamas Almos Vami\corref{onbe}}

\cortext[onbe]{On behalf of the CMS Collaboration}

\affiliation{organization={University of California Santa Barbara},
    addressline={}, 
    city={Santa Barbara},
    postcode={93117}, 
    state={CA},
    country={USA}
}

\begin{abstract}
Long-lived charged particles are predicted by various theories beyond the Standard Model, leading to unique signatures that could reveal new physics. At the LHC, the CMS detector enables searches for these massive particles, identifiable by their characteristic ionization patterns. Using data collected during 2017-2018, we search for signals of anomalous ionization in the silicon tracker. We present a novel approach to background prediction, utilizing the distinct ionization measurements of the silicon pixel and strip detectors as independent variables. We interpret the results within several models including those with staus, stops, gluinos, and multiply charged particles as well as a new model with decays from a Z$'$ boson.
\end{abstract}




\end{frontmatter}


\newcommand{\mumS}{$\mu \text{m}^2\,$}
\newcommand{\ttbar}{$\mathrm{t}\bar{\mathrm{t}}$\,}
\newcommand{\rhadrons}{R-hadrons}
\newcommand{\rhadron}{R-hadron}
\newcommand{\pt}{\ensuremath{p_{\text{T}}}}
\newcommand{\gstrip}{\ensuremath{G_{\text{i}}^{\text{Strips}}}}
\newcommand{\iasp}{\ensuremath{I_{as}^\prime}\,}
\newcommand{\fpix}{\ensuremath{F_{\text{i}}^{\text{Pixels}}}}
\newcommand{\ih}{\ensuremath{I_{\text{h}}}}
\newcommand{\Ih}{\ensuremath{I_{\text{h}}}}
\newcommand{\Irel}{\ensuremath{I_{\text{PF}}^{\text{rel}}}}
\newcommand{\pthscp}{\ensuremath{\pt^{\text{HSCP}}}}
\newcommand{\DeltaRmini}{\ensuremath{\Delta R_{\text{mini-iso}}}}
\newcommand{\Itrk}{\ensuremath{I_{\text{trk}}}}
\newcommand{\dedx}{\ensuremath{\frac{\mathrm{d}E}{\mathrm{d}x}}\,}
\newcommand{\Cgrav}{\ensuremath{C_\text{grav}}\,}
\newcommand{\sigptoverptsq}{\ensuremath{\sigma_{\pt}/\pt^2}\,}
\newcommand{\sigmaptoverpt}{\ensuremath{\sigma_{\pt}/\pt}\,}
\newcommand{\RPF}{\ensuremath{R_{\text{P/F}}}\,}
\newcommand{\RPFj}{\ensuremath{R_{\text{P/F}}(j)}\,}
\newcommand{\Iconetrk}{\ensuremath{I_{\text{trk}}^{\Delta R< 0.3}}\,}
\newcommand{\dz}{\ensuremath{d_{\mathrm{z}}}\,}
\newcommand{\dxy}{\ensuremath{d_{\mathrm{xy}}}\,}
\newcommand{\NFAIL}{\ensuremath{N_{\text{FAIL}}^{\text{bkg}}(j)}}
\newcommand{\NPASS}{\ensuremath{N_{\text{PASS}}^{\text{bkg}}(j)}}

\newpage
\section{Introduction}

The Standard Model (SM) is a quantum field theory that explains the universe at its fundamental level. Often we describe it by its particle content. The force-carrying particles, known as bosons, mediate fundamental interactions: the photon for electromagnetic interaction, gluons for the strong interaction, and the W and Z bosons for the weak interaction. The Higgs boson, a scalar particle, gives mass to elementary particles through the Higgs mechanism.  The remaining particles, called fermions, make up matter and are divided into quarks and leptons. The quarks interact with all the three forces, while the (charged) leptons interact with (electromagnetic and) weak interaction(s). There are three fermion families in the SM.

Despite its many successes, the Standard Model cannot be the ultimate theory of the universe, as it leaves several fundamental questions unanswered. For example, according to the Big Bang theory, equal amounts of matter and antimatter should have been created, yet the observable universe consists almost entirely of matter. What causes this matter-antimatter asymmetry? There is strong evidence of the existence of dark matter, but what the SM does not provide any candidate particle for it. Another puzzle is the hierarchy problem: why is it that the gravity is orders of magnitude weaker than any other force in the SM? 

There has been several theoretical attempts to answer these questions. One of the first popular solution was to introduce superpartners to the SM particles~\cite{Bauer:2009cc,Kusenko:1997si}: each boson would have a fermion partner, and each fermion would have a boson partner. This model is called Supersymmetry (SUSY) and would answer all the questions above. The partner of the quarks and gluons are the squarks and gluinos, while the partner of the leptons are the sleptons. For example, the SUSY partner of the top quark is the stop squark, and the partner of the tau lepton is the stau slepton. Further theories that answer part of the questions above have been proposed, too. Among them the solution is provided by introducing additional heavy Z$'$ bosons, or a 4th generation of fermions ($\tau'$), or several other Higgs bosons.  

\begin{figure}[h!]
   \centering
       \includegraphics[width=.92\textwidth]{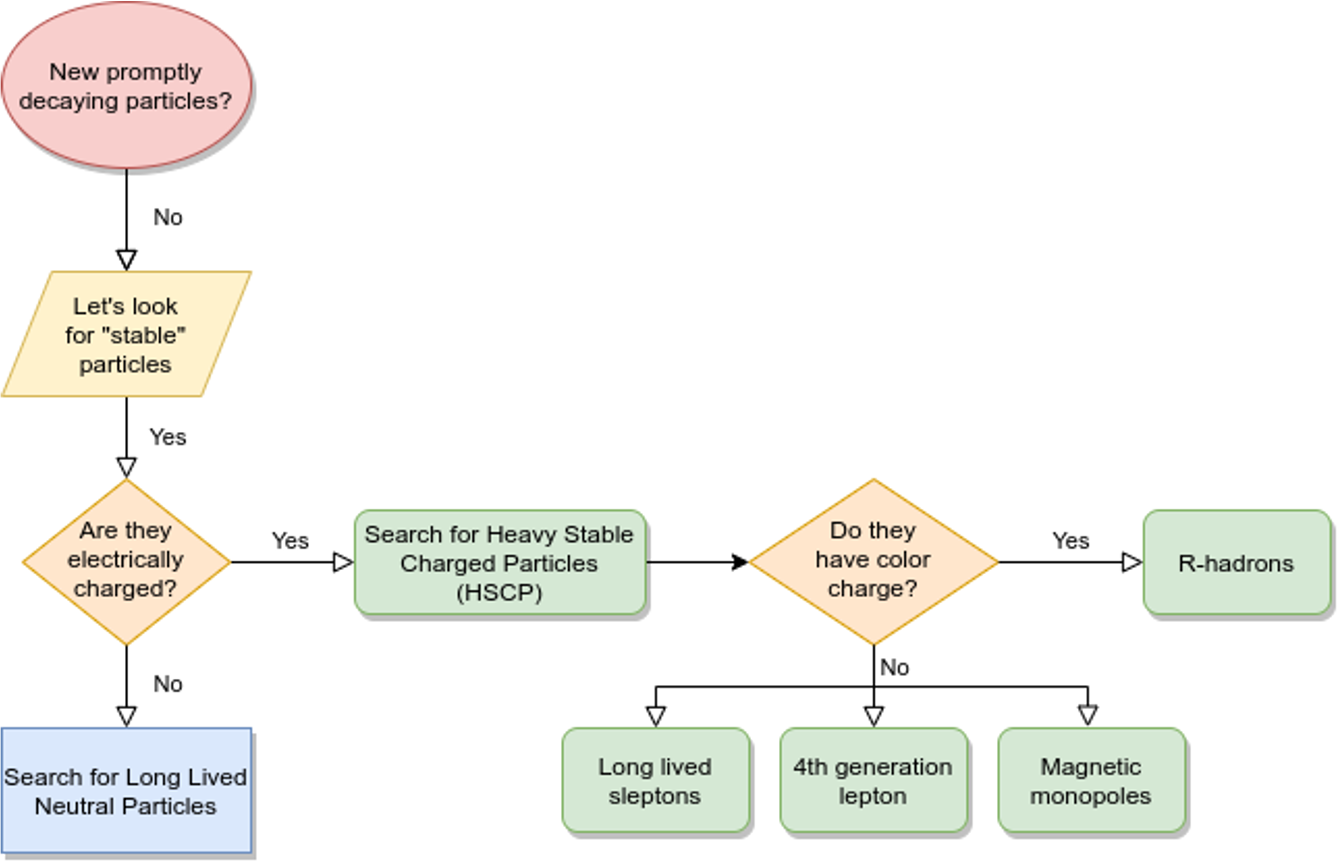}
	\caption{Flow chart for possible new directions in searches.}
   	\label{fig:flowchart}
\end{figure}

These questions and theories have remained open for decades, since no significant experimental evidence for physics beyond the Standard Model (BSM) has been observed to date. The collider searches at the Large Hadron Collider (LHC) and other laboratories have been looking for signatures in which the BSM particles decay promptly. However, there are several mechanisms that can lead to longer lifetime, even in the SM. With this we have a new direction to explore BSM physics and to look for long-lived particles~\cite{Lee:2018pag}. In this work, we search for "stable" particles--meaning not necessarily fundamentally stable, but stable on the detector level, where they traverse the detector without decaying. Figure~\ref{fig:flowchart} shows a flow chart to categorize how these particles interact with different forces and what models would predict their existence. If they are electrically charged, these particles are often referred to as Heavy Stable Charged Particles (HSCPs)~\cite{Drees:1990yw,Fairbairn:2006gg,Koch:2007um,Fargion:2005ep}.

If the HSCPs have color charge that means they will hadronize. A BSM example for this are the \rhadrons\ from split-SUSY~\cite{Giudice:2004tc,ARKANIHAMED20053, Hewett_2004, Kilian_2005}. \rhadrons\ are a bound state of SM and SUSY  particles. Some of these form negative, neutral, positive, or even double positive hadrons. The \rhadron's interactions with matter is described in the model in Ref.~\cite{Kraan:2004tz, Mackeprang:2006gx}. If the HSCPs are colorless, one can think of several theories that would predict them. For example a magnetic monopole~\cite{Schwinger:1966nj}, i.e. the magnetic equivalent of an isolated electric charge, would act as an HSCP. However, it would bend differently, and because of that we do not consider it in this work. Sleptons in GGM SUSY~\cite{Meade_2009}  are predicted to be long-lived. We consider both pair production and cascade production of staus from GMSB SUSY~\cite{Giudice:1998bp,Allanach:2002nj} in this work. Furthermore, 4th generation leptons ($\tau'$) are considered too, both with singly and doubly charge. We consider both pair-production and a production through a boosted Z$'$ boson. This last model is especially interesting, as it was proposed as a solution an excess seen by ATLAS~\cite{Giudice_2022}. 

ATLAS reported a 3 sigma excess with SM compatible v/c ($\beta$)~\cite{ATLAS_excess}. The resolution of this by the phenomenologist was to produce the doubly charged $\tau'$ from the decay of a boosted Z$'$ particle. In this scenario the ionization will be high while the $\beta$ can be high too.

One common aspect of all these models is that they have a distinctively different characteristic ionization loss (dE/dx) from the SM particles. That is why this work presents a signature driven, model independent search with many possible interpretations. The HSCP signature is an isolated track of high $p_T$ with large energy deposits in the tracker (large dE/dx). 

This analysis is carried out using the CMS detector at the LHC. A detailed description of the CMS detector, together with a definition of relevant kinematic variables, can be found in Ref.~\cite{CMS:2008xjf}. All figures in this proceedings are part of the results submitted to JHEP with ArXiv ID of 2410.09164 \cite{cmscollaboration2024searchheavylonglivedcharged}, with supplementary material under CMS-EXO-18-002, also available under HEPData~\cite{hepdata}.

\section{Ionization observables}

We use four observables in this analysis. Eq.~\ref{eq:fpix} shows a discriminator in the pixel detector, called \fpix, where $P'_j$ is a hit level minimum ionizing particle (MIP) compatibility based on the Tracker Detector Performance Group's detailed calibrations,
n is the number of pixel hits (excluding layer 1). 
\begin{equation}
	\fpix = 1 -  \prod_{j=1}^n P'_j \sum_{k=0}^{n-1} \frac{[-\ln( \prod_{j=1}^n P'_j)]^k}{k!}
	\label{eq:fpix}
\end{equation}

The calibrations are based on a very detailed simulation called \textsc{PixelAV}~\cite{Chiochia:2004qh}. These include effects like bias voltage changes, gain calibration changes, temperature, and irradiation dose.
The electric field map input to  \textsc{PixelAV}  is determined using ISE TCAD~\cite{tcad} simulations of a pixel cell. The Lorentz drift vs. substrate depth has been tuned to calibration data. The simulation also includes carrier trapping, diffusion, charge induction on the sensors and noise simulation. 
The \fpix\ templates have been produced 25 times over the 2017--2018 period, corresponding to an update about every 5 fb$^{-1}$.

The second discriminator, using the strip detector, is shown in Eq.~\ref{eq:gstrips}, where $P_j$ is a hit level MIP probability for a charge to be smaller/equal to the measured charged based on our templates calibrated at low momentum (20-48 GeV) data, N is the number of (cleaned) hits in the strips detector.

\begin{equation}
	\gstrip = \frac{3}{N} \, \left(
   \frac{1}{12N} + \sum_{j=1}^N
   \left[
   P_j \, \left( P_j - \frac{2j-1}{2N} \right)^2 \right] \right)
   \label{eq:gstrips}
\end{equation}

\begin{figure}[ht!]
   \centering
       \includegraphics[width=.42\textwidth]{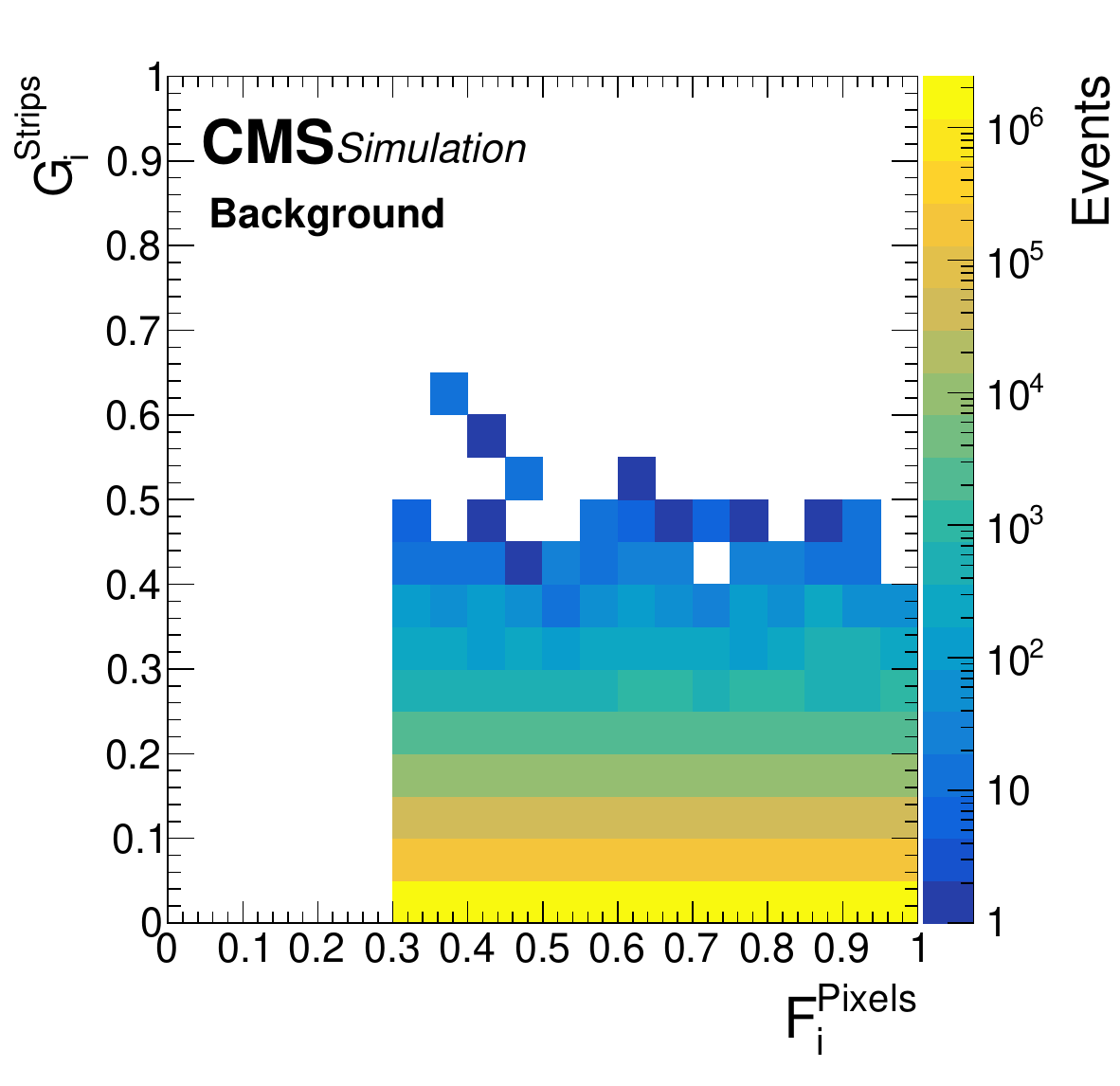}
       \includegraphics[width=.42\textwidth]{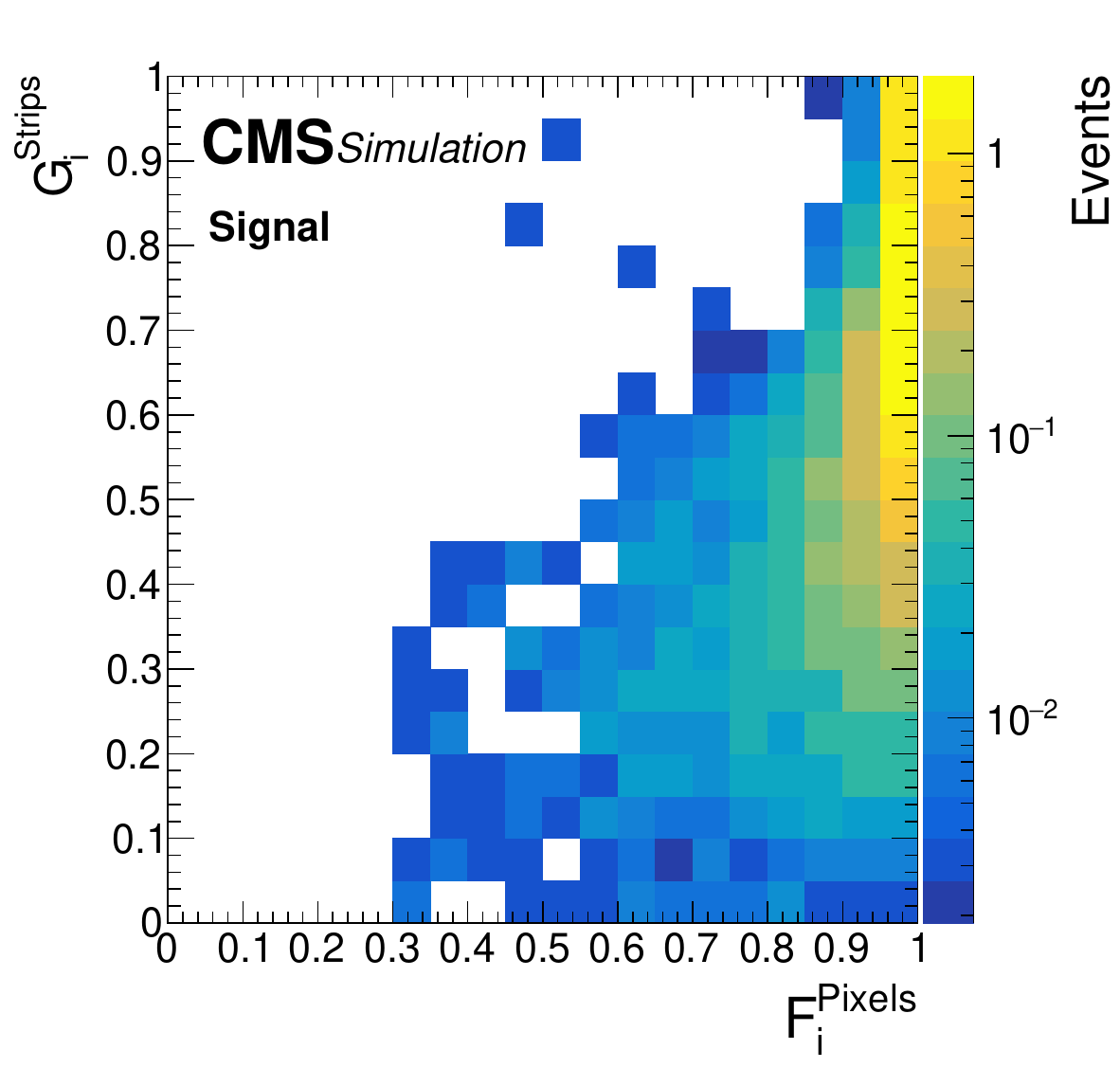}
	\caption{The \fpix\ vs. \gstrip\ distributions for the SM background (QCD, W+jet, \ttbar)  simulation (left), and an 1800~GeV gluino R-hadron simulation (right), for events after pre-selection.}
   \label{fig:GiVsFi}
\end{figure}

These two quantities are uncorrelated by construction. Figure~\ref{fig:GiVsFi} shows the \fpix\ vs. \gstrip\ distributions for both signal (1800 GeV gluino R-hadron) and background (QCD, W+jet, \ttbar) simulations. Indeed we can observed that the two quantities are not correlated for background, while they are accumulated in the high end in both quantities for the signal.

Besides these discriminators, we also use a dE/dx estimator~\cite{Chatrchyan_trk10} as shown in Eq.~\ref{eq:MassFromHarmonicEstimator}. 
\begin{equation}
	\Ih = K\frac{m^2}{p^2}+C
 \label{eq:MassFromHarmonicEstimator}
\end{equation}

This uses a squared harmonic mean in order to suppress the tails of the Landau distribution, which happen quite often. This estimate, on the other hand, is sensitive to lower fluctuations. However, those do not happen very often, one example can be a high voltage change on the sensors.

The mass of the HSCP can be interpreted through approximating the Bethe--Bloch formula with $I_h(m,p,K,C)$, where the empirical parameters $K$ and $C$ are determined using a sample of low-momentum particles composed of protons, kaons and pions. The calibrated value of these parameters are $K=2.5\pm0.01$  MeV/cm and $C=3.14\pm0.01$ MeV/cm.

\begin{figure}[ht!]
    \centering
        \includegraphics[width=.45\textwidth]{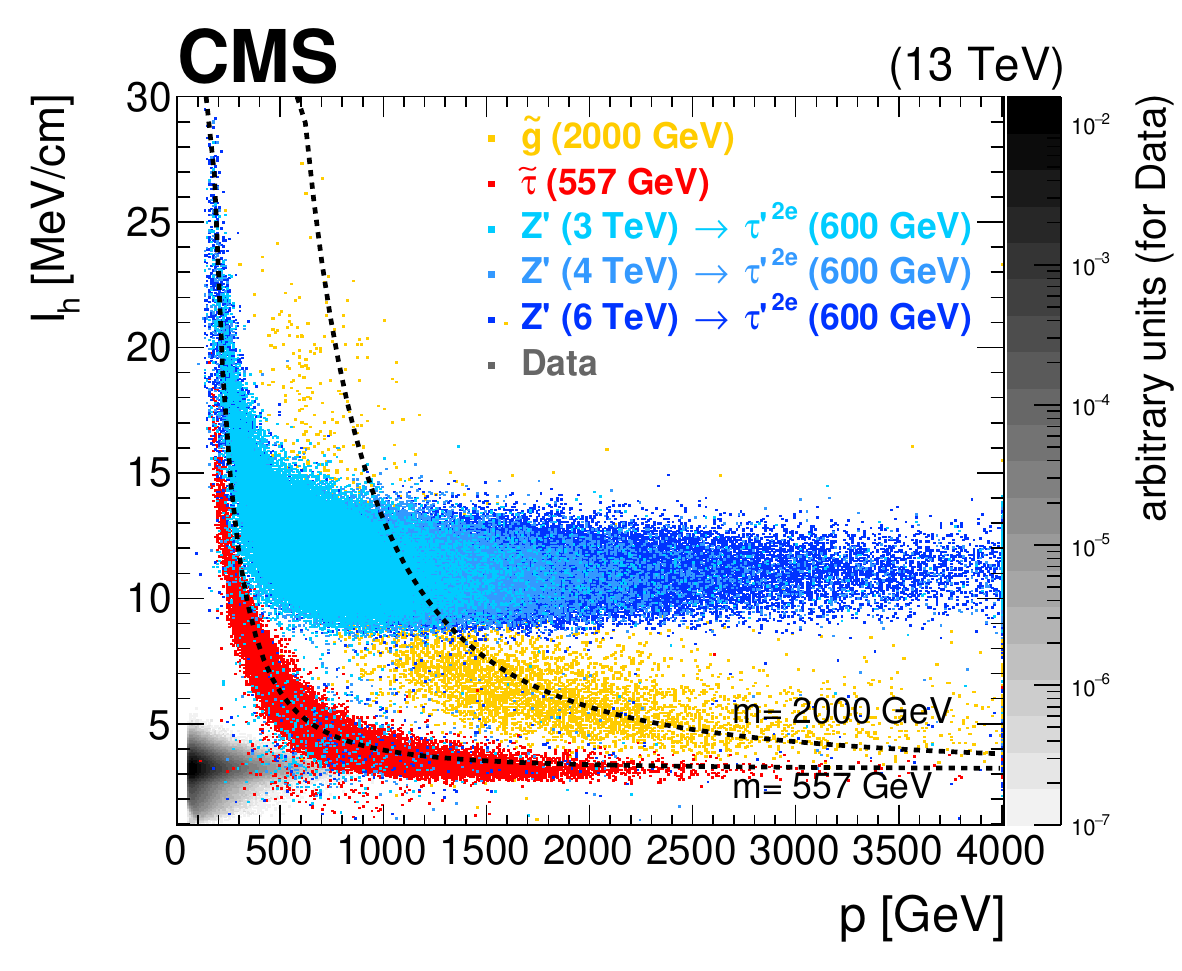}
        \includegraphics[width=.45\textwidth]{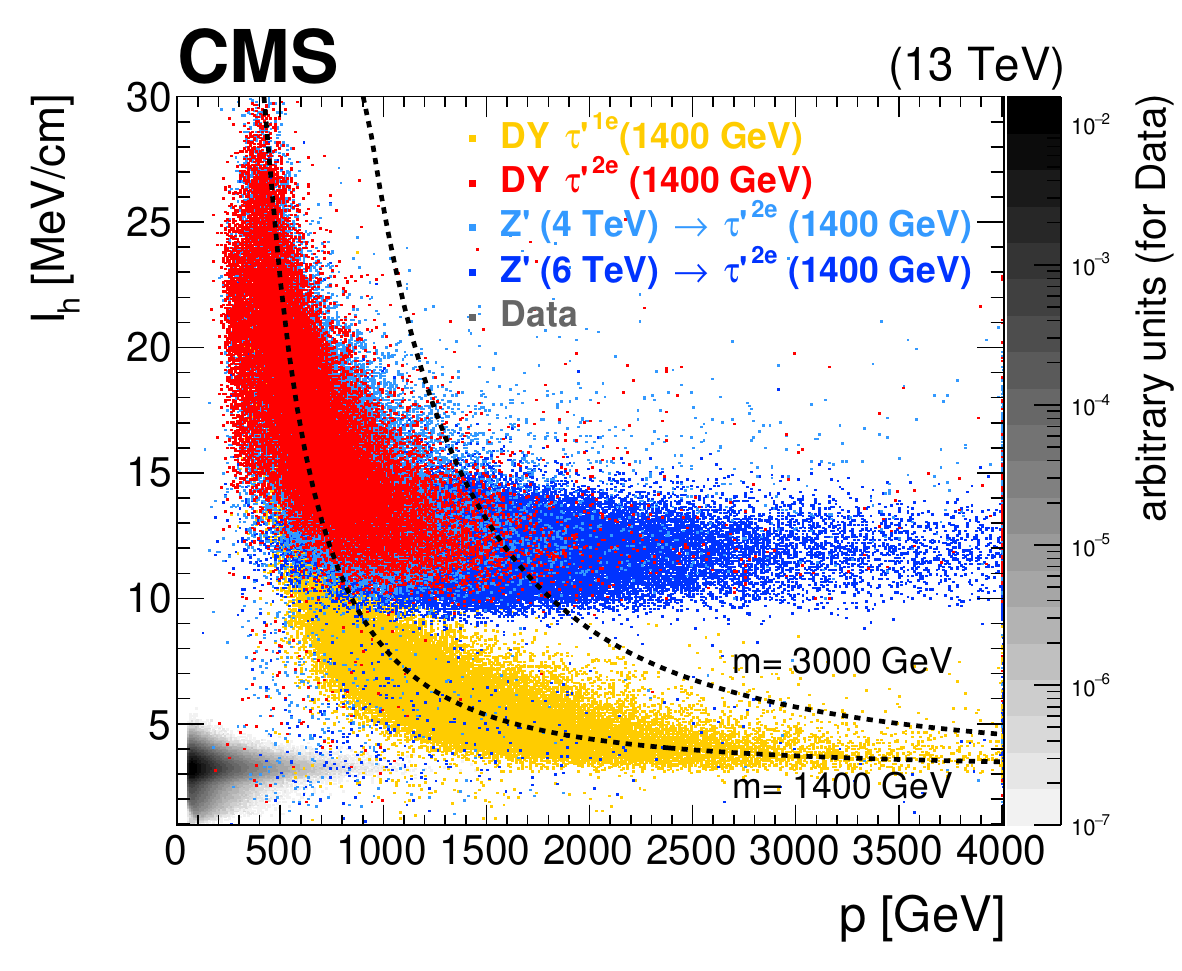}
	 \caption{The distribution of $I_h$ vs. HSCP candidate $p$. Different colors correspond to different signal simulations.
	 Data is shown in a grey histogram, normalized to unit area.
	 Based on Eq.~ \ref{eq:MassFromHarmonicEstimator} we show the distribution for a particle mass of 557 and 2000~GeV on the left, and 1400 and 3000~GeV on the right using the dashed lines.}
    \label{fig:dedx_vs_momentum}
\end{figure}

The distribution of $I_h$ vs. HSCP candidate $p$ can be seen in Fig.~\ref{fig:dedx_vs_momentum} for different signal simulations in colors and for data as a histogram. The dashed lines represent the distribution for a particle with a mass of 557 and 2000~GeV on the left, and 1400 and 3000~GeV on the right. These are based on  Eq.~ \ref{eq:MassFromHarmonicEstimator}.

\section{Understanding the background and preselection}

We used Monte Carlo (MC) SM simulations to understand our background. From these studies we concluded the following list of sources:

\begin{itemize}
    \item Fake tracks
    \item Bad ionization measurement
    \item Being in the tail of the Landau distribution
    \item Overlapping tracks in the tracker coming from
    \begin{itemize}
        \item Pileup
        \item Boosted mesons decays to overlapping dileptons
        \item Tracks in the core of jets
    \end{itemize}
\end{itemize}

\begin{figure}[ht!]
    \centering
        \includegraphics[width=.7\textwidth]{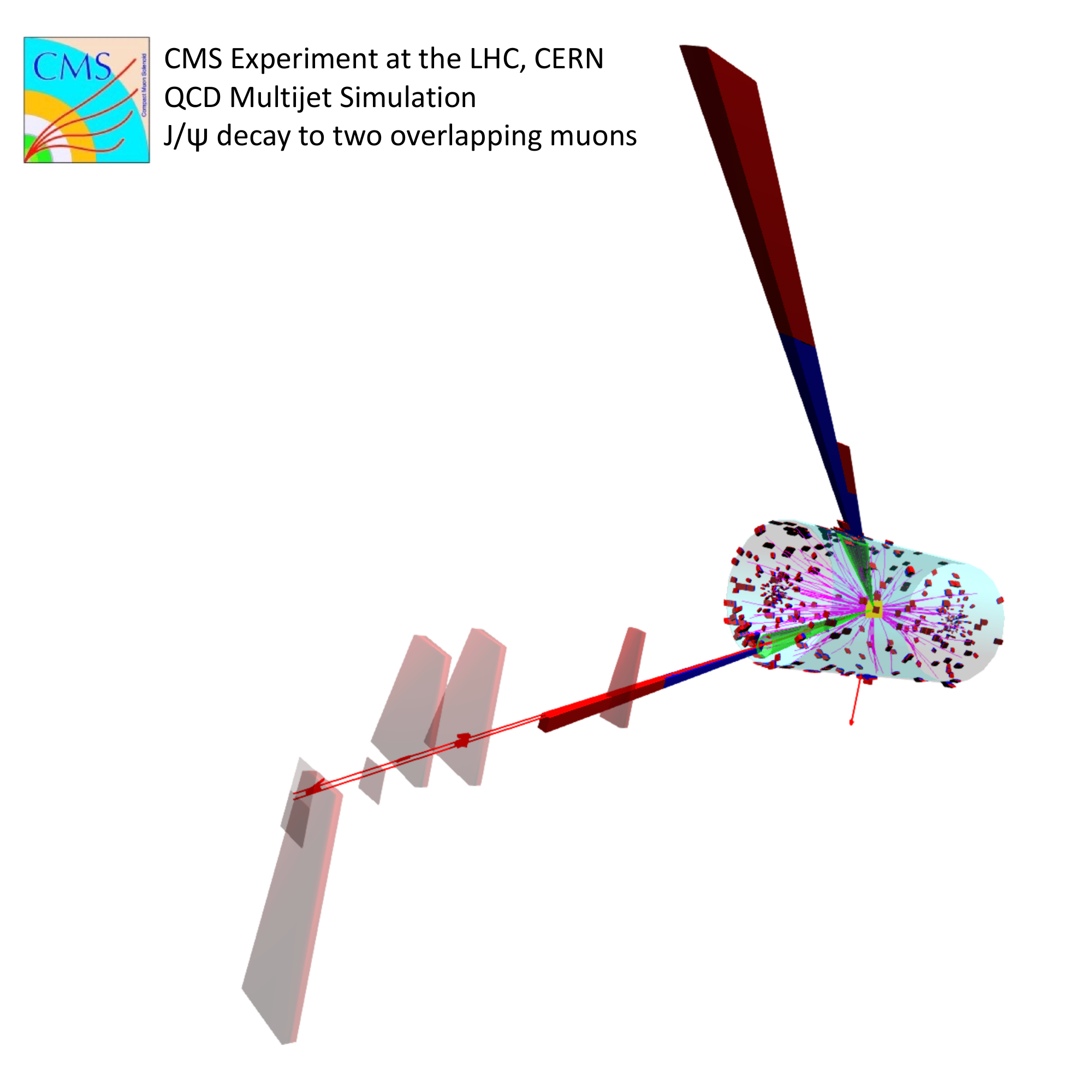}
	 \caption{Event display of a boosted  J/$\Psi$ particle decaying into two overlapping muons.}
    \label{fig:EventDisplayJPsi}
\end{figure}

Fig.~\ref{fig:EventDisplayJPsi} shows an example in MC where a boosted J/$\Psi$ particle is created and decays to two muons, that are overlapping in the tracker and they are also very close in the muon chambers.

With this understanding we optimize our preselection. We used 2017--2018 data corresponding to 101~fb$^{-1}$ data. Exclusion of 2016 data is motivated by the read-out chip issues in the strips detector and having the Phase-0 pixel detector~\cite{Phase1Pixel}. The cutflow after applying each preselection cut is shown in Table~\ref{tab:cutflow}. The details of these tracking criteria are explained in Ref.~\cite{Chatrchyan:1704291}. 

\begin{table}[ht]
	\centering
	   \begin{tabular}{|l|c|c|c|}
       \hline
Selection criteria & Data & gluino (1.8~TeV) &  stau (557~GeV) \\
\hline
All events &    1  &   1  &   1  \\
\hline
Trigger &  0.15  & 0.11  & 0.86  \\
\hline
$\pt > 55~GeV$ &  0.11  & 0.11  & 0.86  \\
\hline
$|\eta| < 1$ &  0.059  & 0.074  & 0.64  \\
\hline
\# of valid pixel hits in L2-L4 $\geq 2$  &  0.056  & 0.071  & 0.62  \\
\hline
Fraction of valid hits  $> 0.8$  &  0.052  & 0.069  & 0.62  \\
\hline
\# of dE/dx measurements $\geq 10$ &  0.052  & 0.069  & 0.62  \\
\hline
High-purity track &  0.052  & 0.069  & 0.62  \\
Track $\chi^2$/dof $< 5$  &  0.052  & 0.069  & 0.62  \\
\hline
$\dz < 0.1$~{cm}  &  0.052  & 0.069  & 0.62  \\
\hline
$\dxy < 0.02$~{cm} &  0.048  & 0.069  & 0.62  \\
\hline
$\Irel < 0.02 $ &  0.014  & 0.065  & 0.61  \\
\hline
$\Itrk  < 15~GeV$ &  0.014  & 0.065  & 0.61  \\
\hline
PF $E/p < 0.3$ &  0.014  & 0.064  & 0.61  \\
\hline
\sigptoverptsq $< 0.0008$ &  0.014  & 0.064  & 0.61  \\
\hline
\fpix $> 0.3$ &  0.011  & 0.064  & 0.60  \\
\hline
  \end{tabular}
  	 \caption{Cumulative selection efficiency for the data and for two signal hypotheses.}
	 \label{tab:cutflow}
\end{table}

The definition of the relative isolation~\cite{Rehermann_2011} is in Eq.~\ref{eq:Irel}.:

\begin{equation}
	\Irel = \frac{(\sum_{\Delta R<\DeltaRmini} \pt^{\text{PF}})-\pthscp} {\pthscp},
	\label{eq:Irel}
\end{equation}

where
\begin{equation*}
	\DeltaRmini = \begin{cases}
         0.2, & \text{ $\pthscp < 50~GeV$} \\
		10~GeV/\pthscp, & \text{ $50~GeV < \pthscp < 200~GeV$ } \\
		0.05, & \text{ $\pthscp > 200~GeV$}.\\
    \end{cases}
\end{equation*}

The definition of the tracker-based isolation is shown in  Eq.~\ref{eq:tkrel}. 

\begin{equation}
	\Itrk = \biggl( \sum_{\Delta R<\DeltaRmini}\pt^{\text{trk}} \biggr) -\pthscp
	\label{eq:tkrel}
\end{equation}

Preselection cuts are ensuring a clean track environment, homogeneous sensor technology, they remove damaged clusters, pileup, and overlaying tracks, e.g. from boosted meson decays (J/$\Psi$) as mentioned earlier.

For the trigger choice we decided to use a single muon trigger. This is motivated by the ATLAS excess being compatible with muons and also leads to less QCD and more electroweak processes as background.

\begin{figure}[ht!]
    \centering
        \includegraphics[width=.45\textwidth]{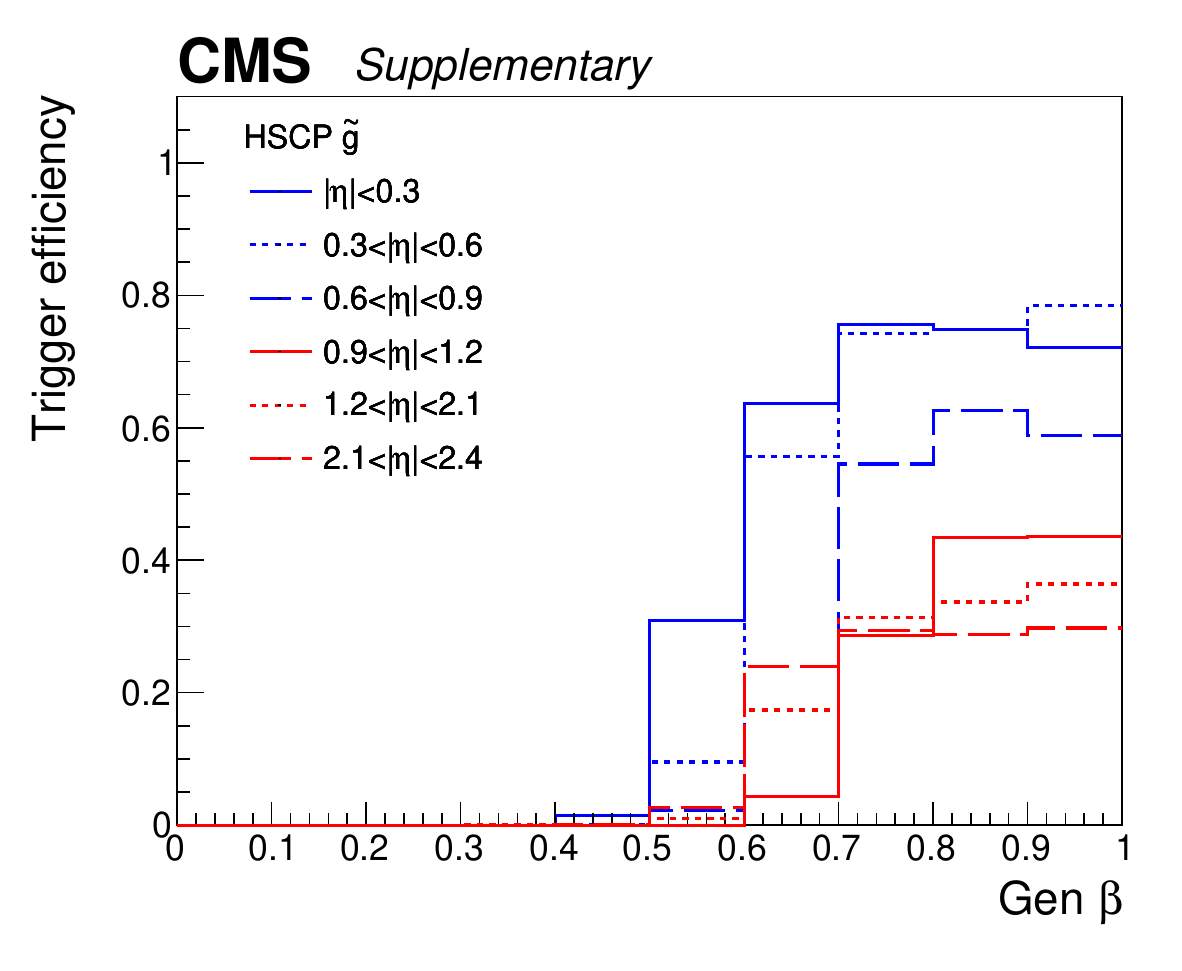}
        \includegraphics[width=.45\textwidth]{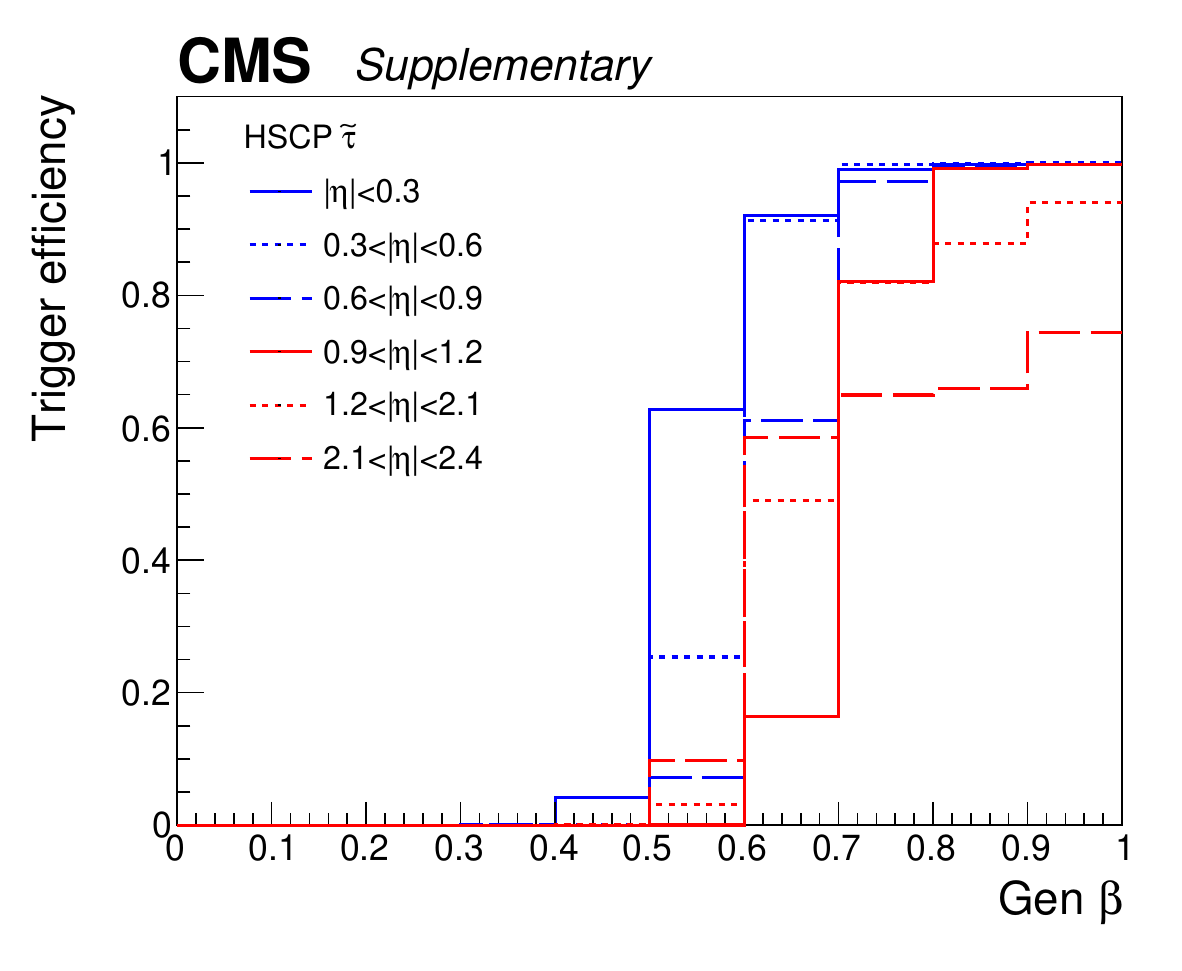}
	 \caption{Trigger efficiency as a function of generated $\beta$ for gluino (left) and stau (right) in bins of pseudorapidity.}
    \label{fig:trigEff}
\end{figure}

Figure~\ref{fig:trigEff} shows the trigger efficiency binned in pseudorapidity ranges for gluinos (left) and staus (right). We can clearly see that this trigger choice is more efficient for muon-like signal like the staus and the $\tau'$ particles. However, we still are able to trigger on the \rhadrons, too. We can also observed that the barrel detector, displayed with blue curves, has higher efficiency.

\section{Data-driven background prediction}

We use two independent methods for background prediction in this analysis, which together provide an additional handle, especially in light of the ATLAS excess. Both of them are data driven, and reuse the trigger selection, preselection, and signal systematics. The first method is referred as the ionization method. This is a new approach, relying on independence of the pixel detector and strip detector~\cite{Chatrchyan_trk10,Chatrchyan:1704291}, it uses a transfer function and invokes a shape based analysis. The second one is the mass method. This is the improved historical approach~\cite{Khachatryan:2011ts,Chatrchyan:2012sp,Khachatryan:2016sfv,EXO-12-026}, assuming the independence of $I_h$ and p as well as the \gstrip\ and $p_T$  as a counting experiment in a dedicated mass window. In this proceedings, we concentrates on the ionization method for simplicity.

The ionization method is a fully data driven method relying on the independence of the pixels and strips detectors. The \fpix\ variable spans the PASS (\fpix\ $>$ 0.9) and FAIL regions (0.9 $>$ \fpix\ $>$ 0.3). We perform a simultaneous fit in bins of the \gstrip\ distribution using the full \gstrip\ shape. The  background yield in $j$-th bin of the  \gstrip\ distribution in the PASS region is determined using Eq.~\ref{eq:pass}, where
\NFAIL\ is a freely floating parameter that determines the background yield in the $j$-th bin, which is constrained to values close to the observed data yield during the fit, and 
 \RPFj\ is the value of the pass-to-fail ratio for bin $j$.

 \begin{equation}
	 \NPASS = R_{\text{P/F}}(j) \, \NFAIL
	 \label{eq:pass}
\end{equation}

We used a constant \RPF\ across the bins $j$. Higher-order polynomials were tested using Fisher's $F$-test~\cite{WileyChi2,BAKER1984437} and did not provide significantly better  results as the constant function. We choose the binning of the \gstrip\ distribution so the last bin contains $\mathcal{O}(1)$ event in the FAIL region for the data.

\begin{figure}[ht!]
   \centering
       \includegraphics[width=.42\textwidth]{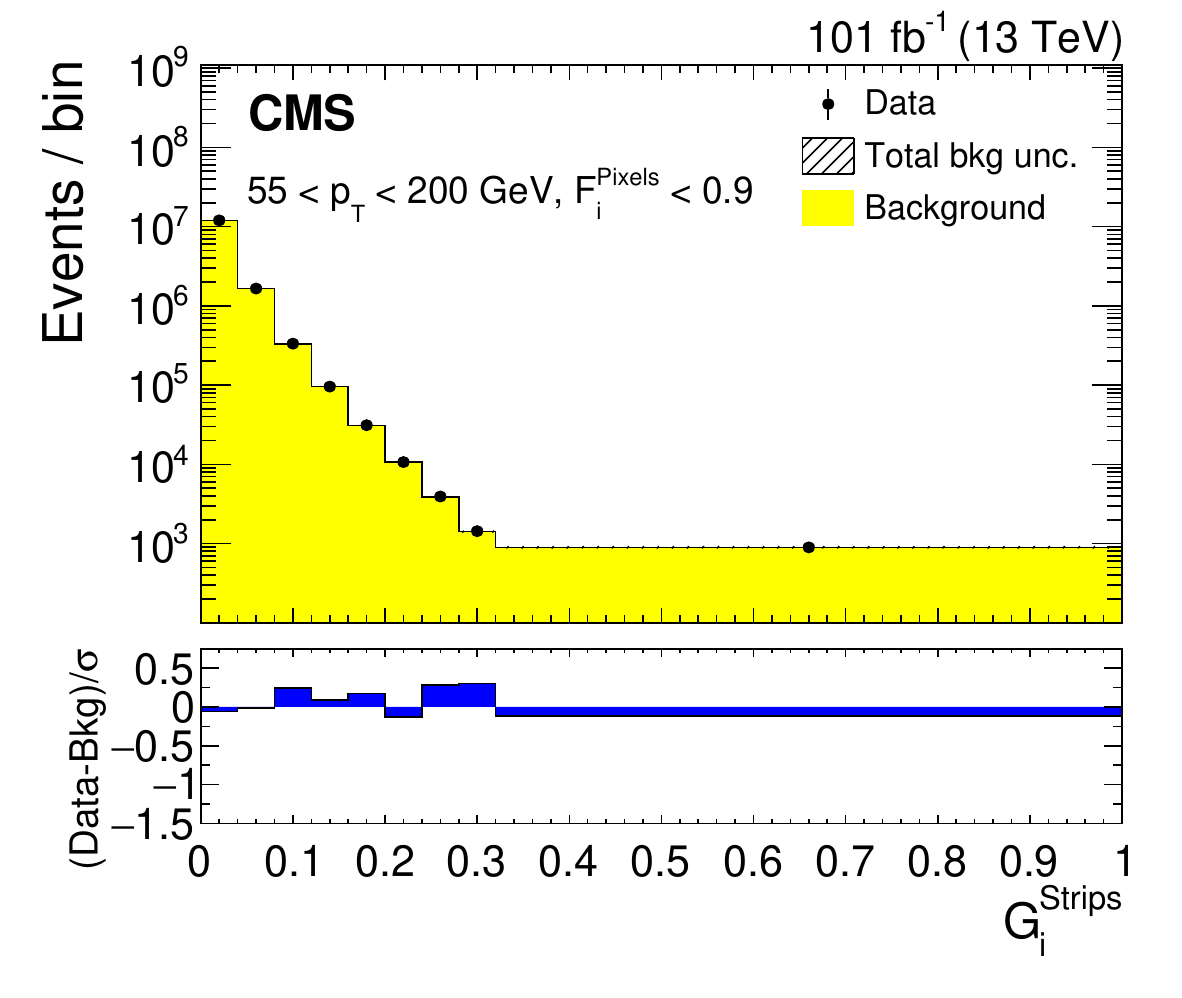}
       \includegraphics[width=.42\textwidth]{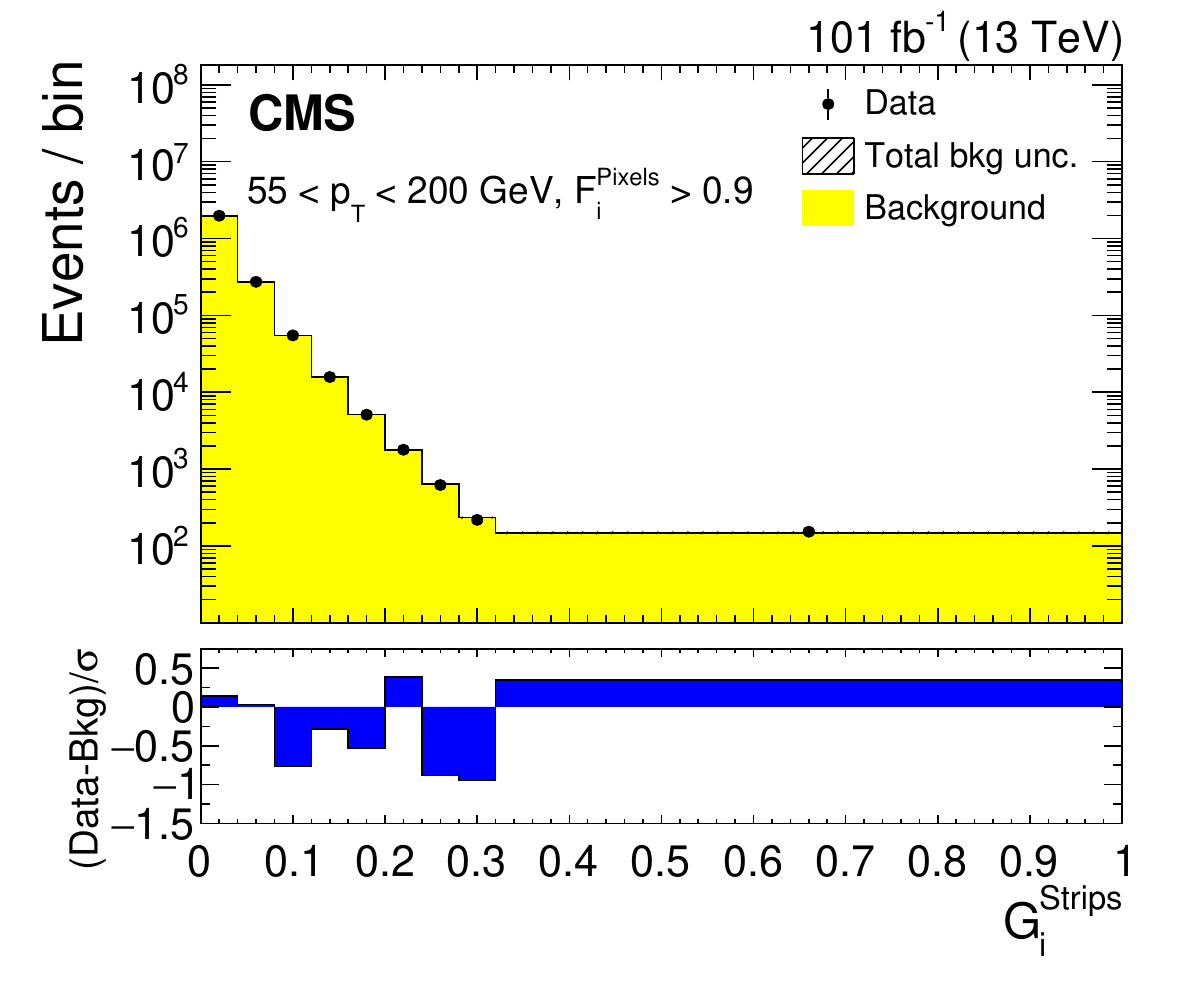}
        \caption{Control region with  55 $<p_{T} <$ 200~GeV, showing the FAIL region in the left and the PASS region in the right.}
   \label{fig:cr}
\end{figure}

We defined two control regions, one with 50 $< p_{T} < $ 55~GeV and 55 $<p_{T} <$ 200~GeV. The background prediction performs very well in these regions, Fig.~\ref{fig:cr} shows the control region for  55 $<p_{T} <$ 200~GeV, left being the FAIL and right the PASS region.

\section{Results}
\begin{figure}[ht!]
   \centering
       \includegraphics[width=.42\textwidth]{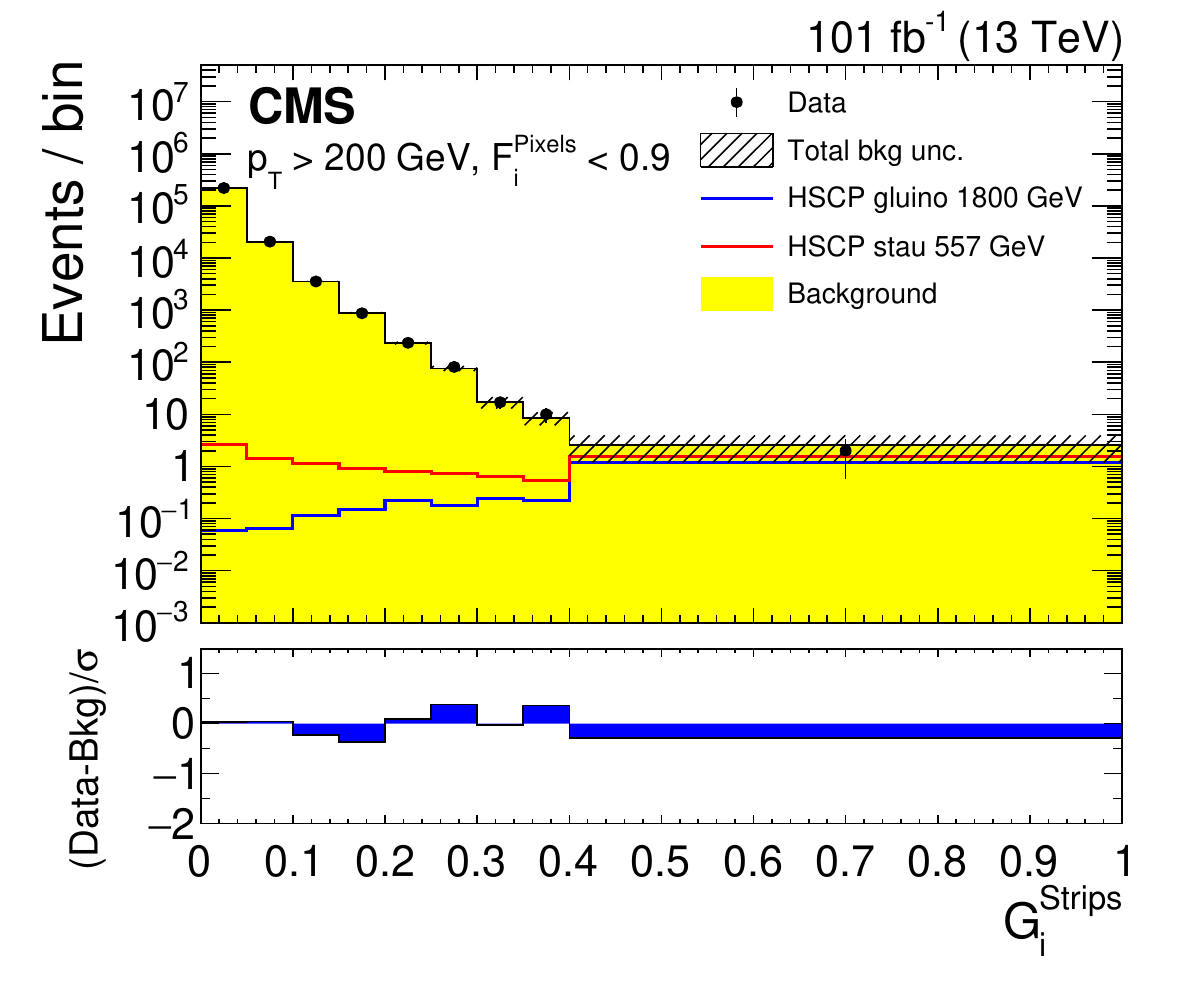}
       \includegraphics[width=.42\textwidth]{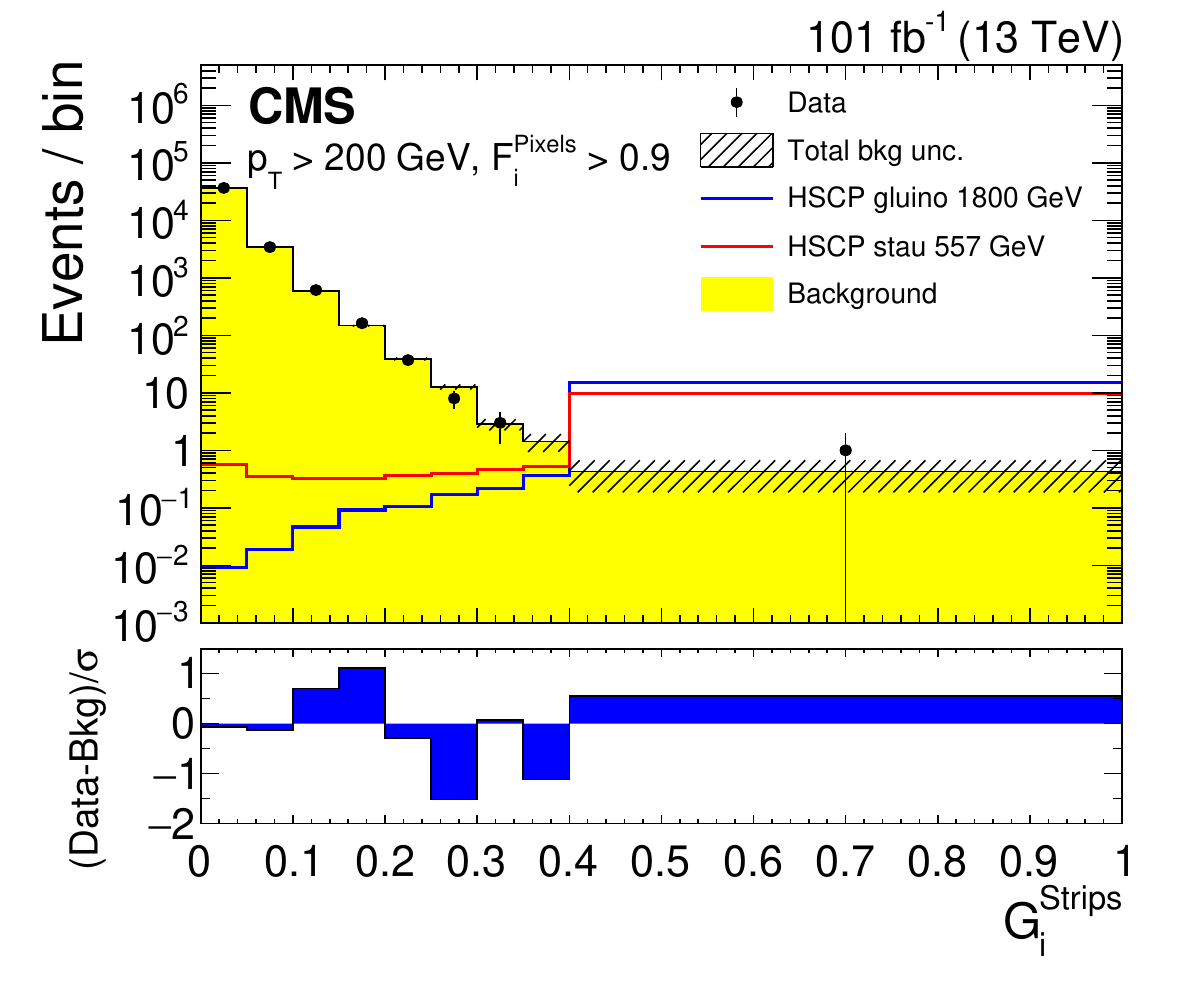}
        \caption{Signal region with  $p_T > $ 200~GeV, \fpix $>$0.9, left is FAIL and right is PASS. The results are consistent with the SM.}
   \label{fig:sr}
\end{figure}

Given that the the background prediction performs well in the control region, we can look at the signal region next. This is defined as $p_T > $ 200~GeV, \fpix $>$0.9 and the full shape of \gstrip is used. Figure~\ref{fig:sr} shows the signal region for FAIL (left) and PASS (right). The results are consistent with the SM. In the last bin we observe one event, where $0.4 \pm 0.2$ were expected. The bin before one event was expected and zero is observed.

\begin{figure}[ht!]
    \centering
        \includegraphics[width=.49\textwidth]{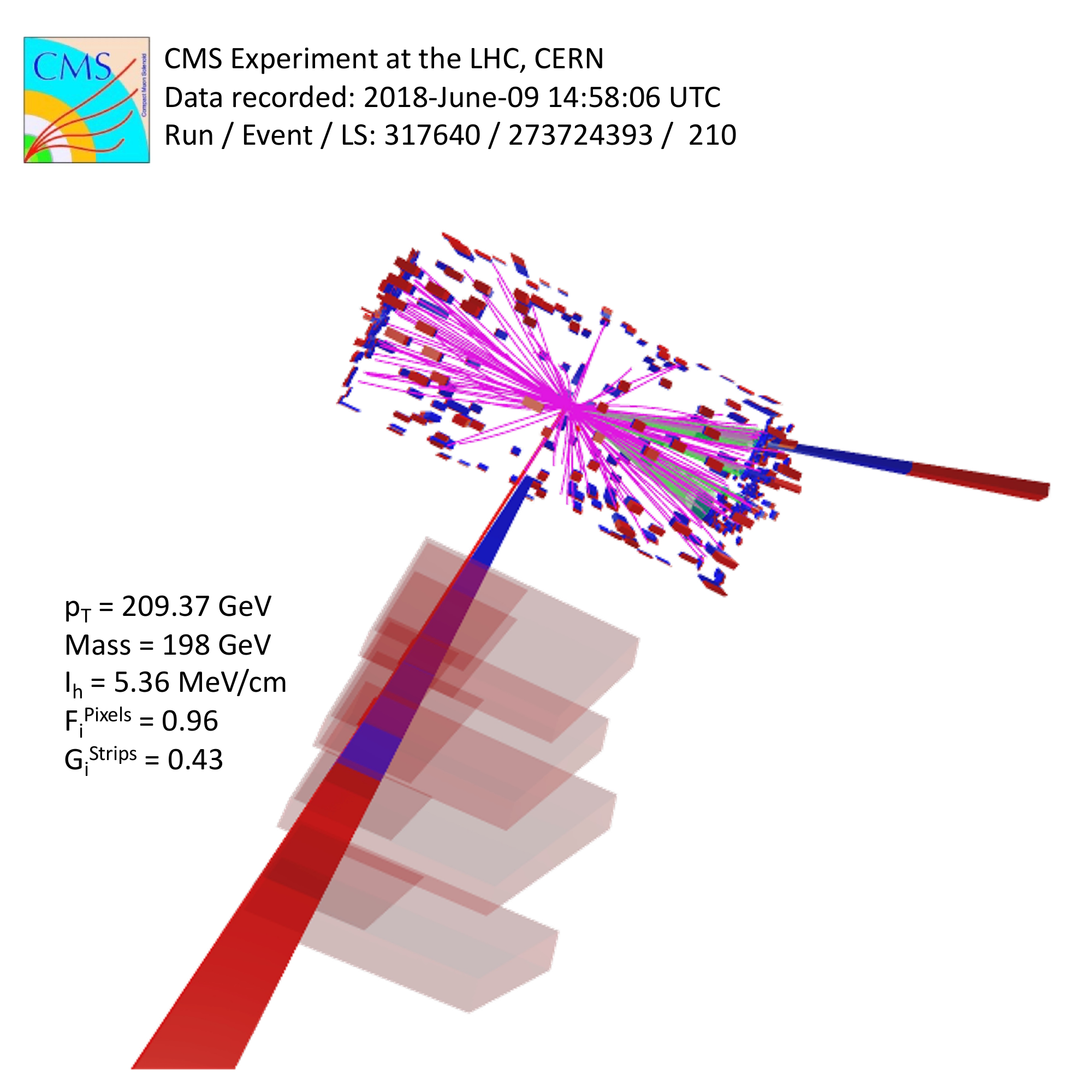}
        \includegraphics[width=.49\textwidth]{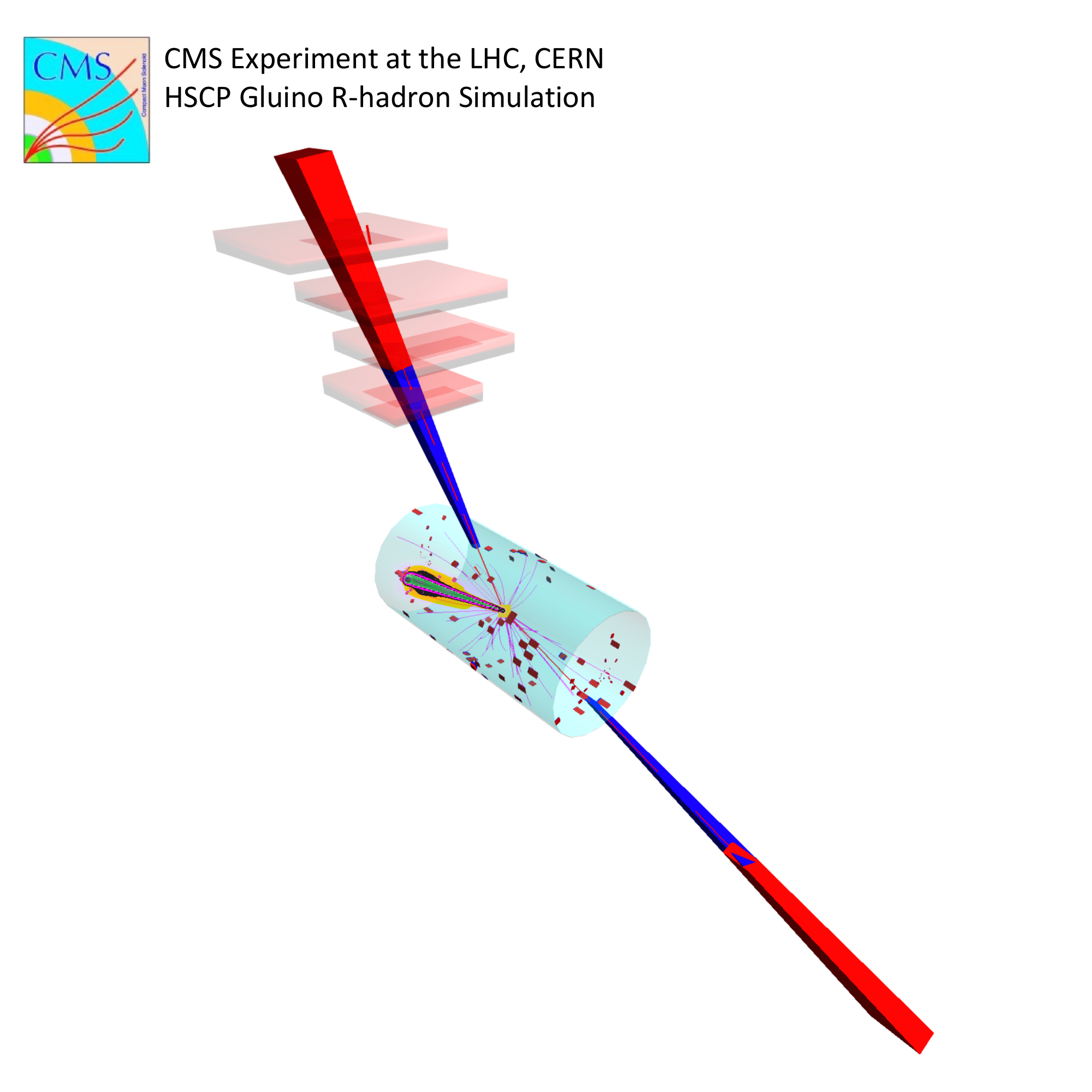}
	 \caption{Visualization of the "excess" event (left) and a simulated gluino R-hadron event (right).}
    \label{fig:EventDisplayData}
\end{figure}

The visualization of the "excess" event above can be seen in Fig.~\ref{fig:EventDisplayData} (left). It has \gstrip = 0.43 and \fpix = 0.96 and the $I_h$ = 5.35 MeV/cm. The mass corresponding to this candidate is 198 GeV. For illustration purposes, we also show a simulated gluino R-hadron event in the right.

Given that no significant excess beyond the SM is observed, we set limits on a variety of signal models. The full CLs method ~\cite{Junk:1999kv, READ:JPG2002} is used as test statistic~\cite{Cowan:2010js} and using a lognormal model for nuisance parameters affecting yields. The CMS \textsc{Combine} tool is used to perform the statistical analysis~\cite{CAT-23-001}.

\begin{figure}[ht!]
   \centering
       \includegraphics[width=.42\textwidth]{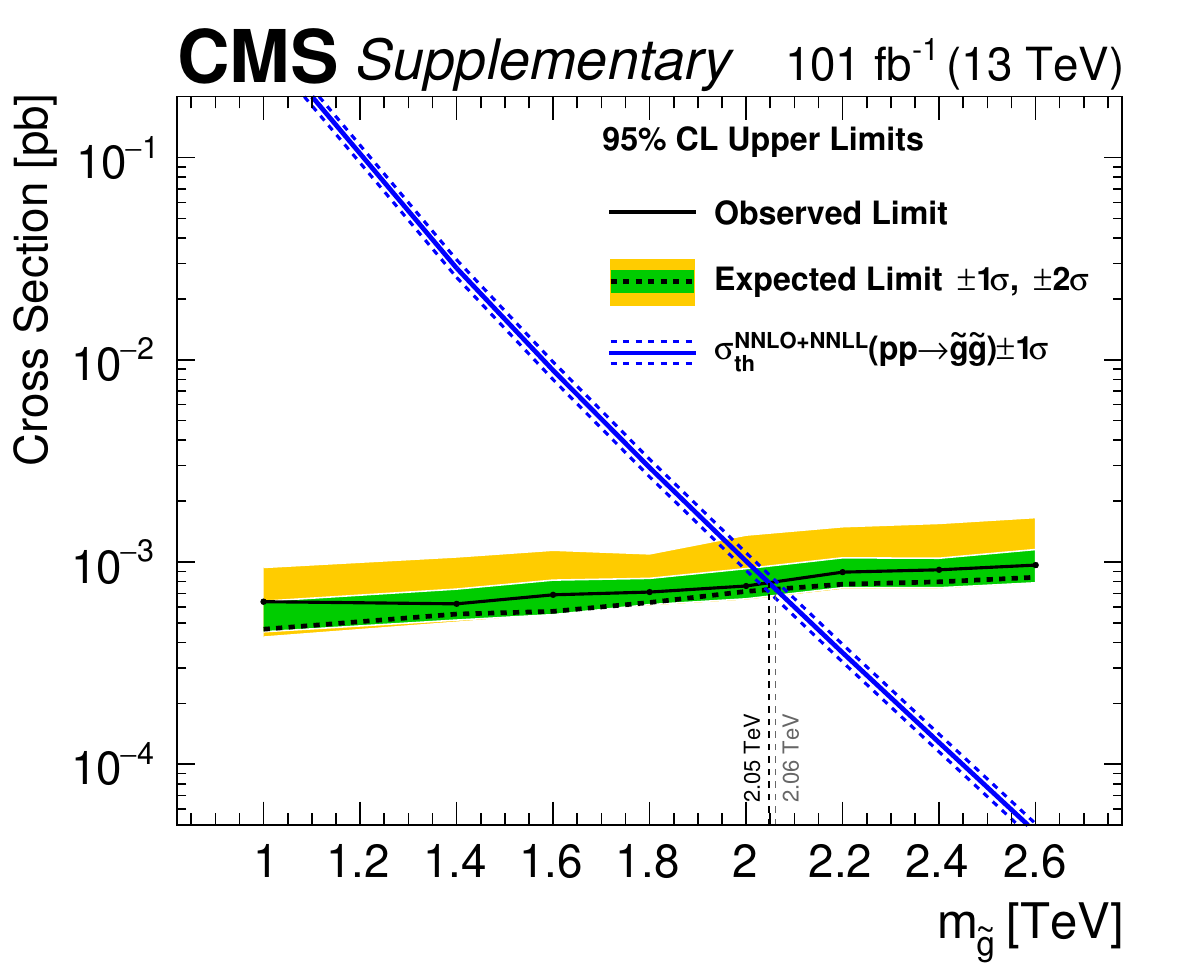}
       \includegraphics[width=.42\textwidth]{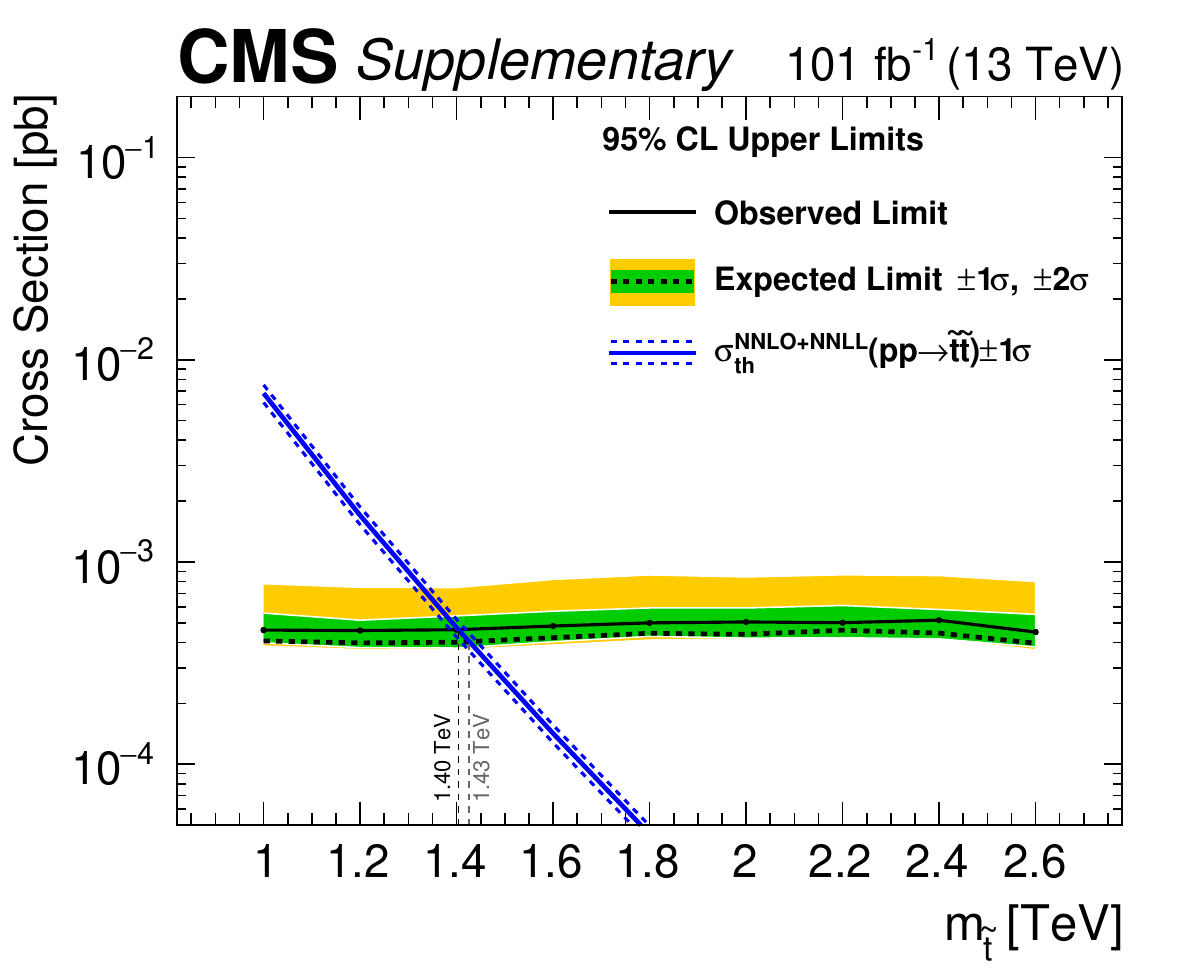}
        \caption{Exclusion limits for gluino R-hadrons (left) and stop R-hadrons (right).}
   \label{fig:limitHad}
\end{figure}

Figure~\ref{fig:limitHad} shows the exclusion limits for gluino R-hadrons (left) and stop R-hadrons (right). The observed (expected) mass limits for ionization method are 2.06 (2.06) TeV for the gluino and 1.40 (1.43) TeV for the stop. The uncertainty bands are assymetric since we are in the low background case where downward fluctuations reach zero already.

\begin{figure}[ht!]
   \centering
       \includegraphics[width=.42\textwidth]{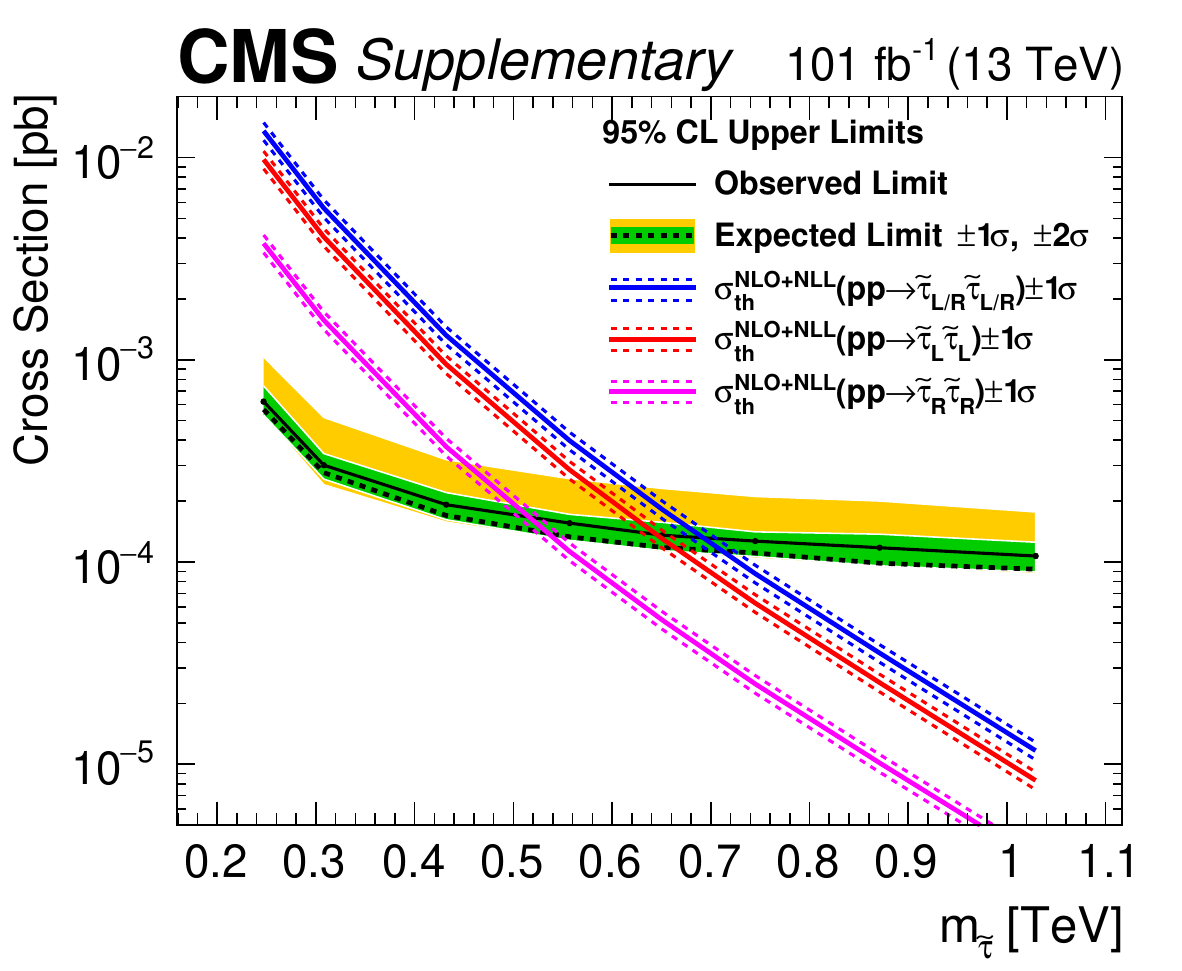}
       \includegraphics[width=.42\textwidth]{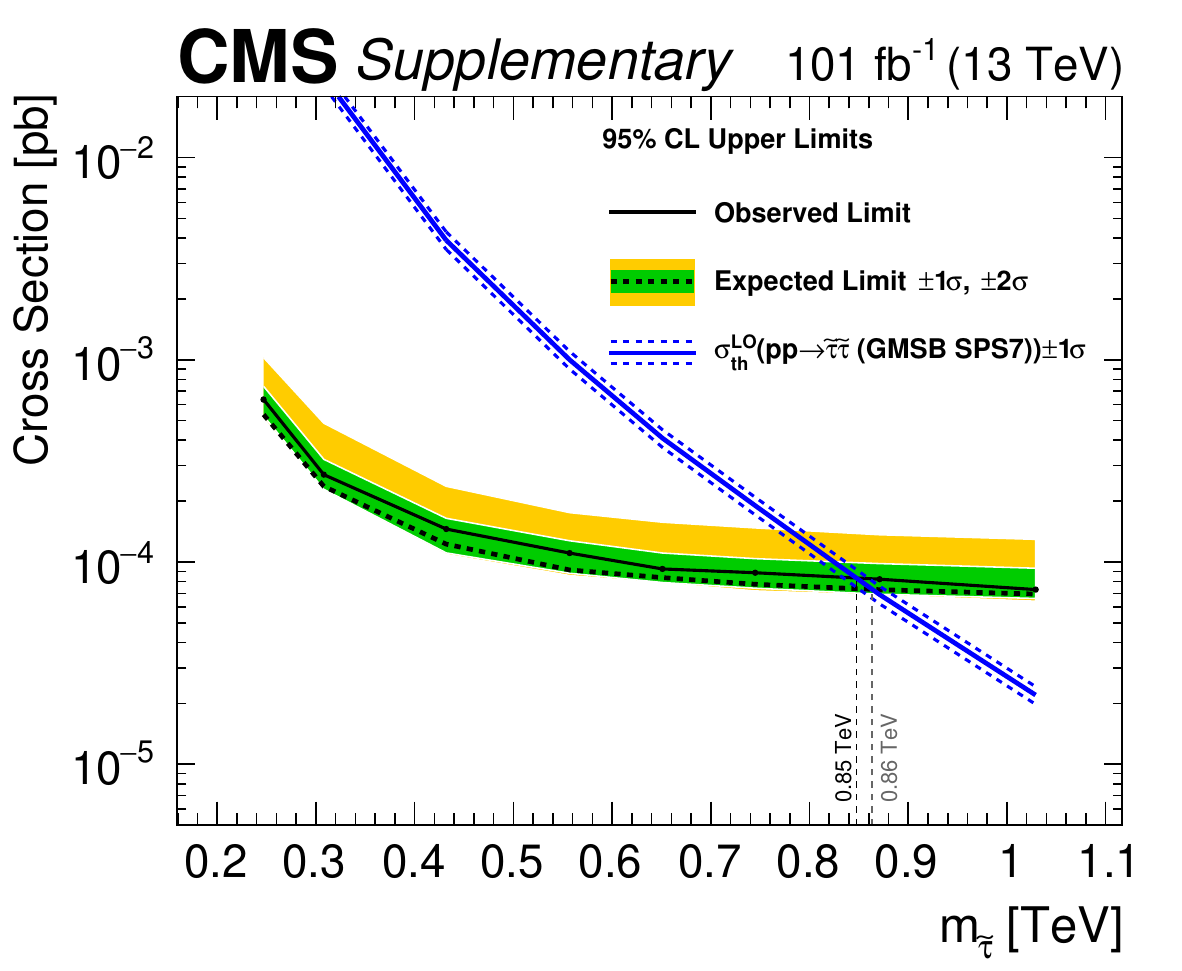}
        \caption{Exclusion limits for pair-produced stau (left) and GMSB produced stau (right).}
   \label{fig:limitLep}
\end{figure}

Figure~\ref{fig:limitLep} shows the exclusion limits for pair-produced stau with different handedness (left) and for GMSB produced stau (right). 

\begin{figure}[ht!]
   \centering
       \includegraphics[width=.42\textwidth]{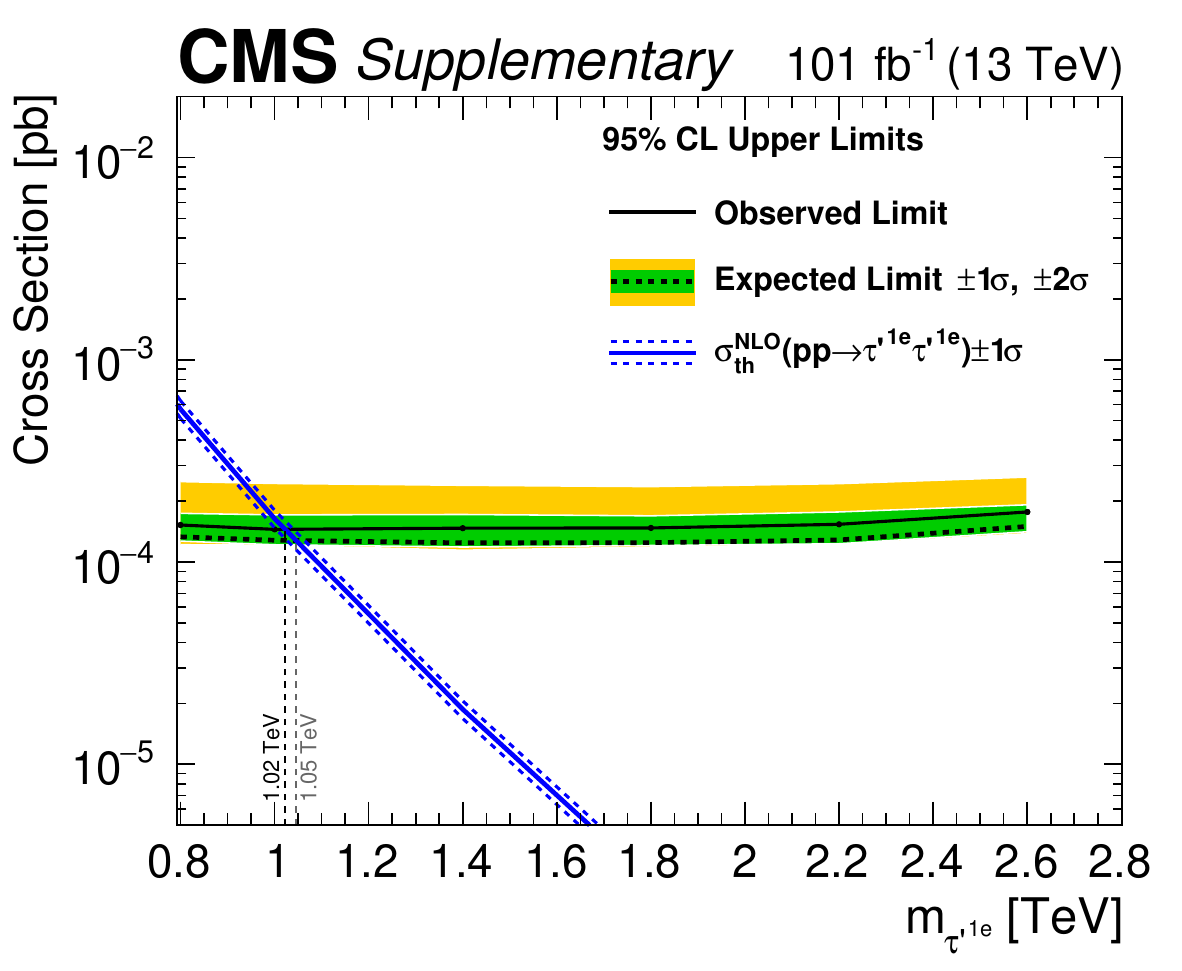}
       \includegraphics[width=.42\textwidth]{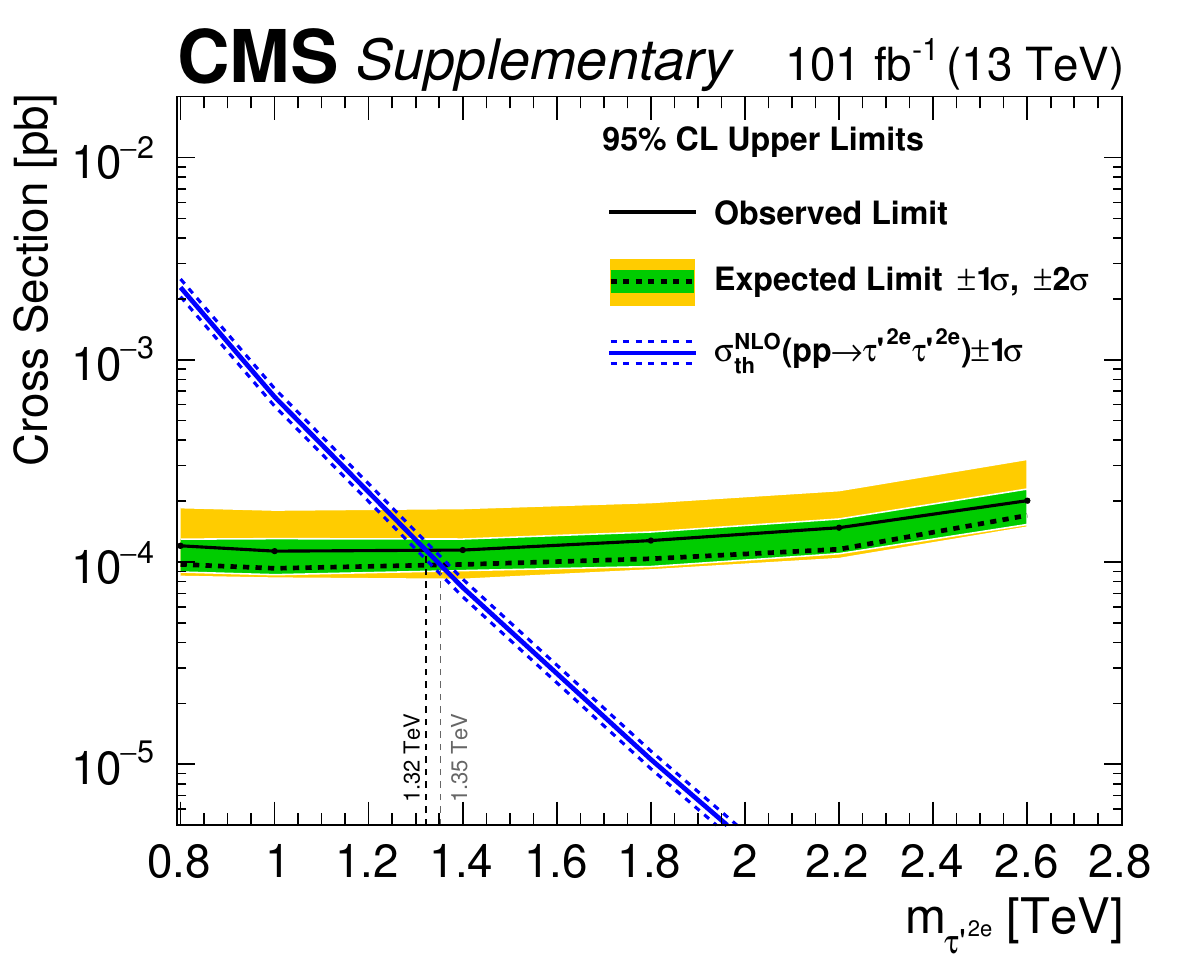}
        \caption{Exclusion limits for pair-produced stau (left) and GMSB produced stau (right).}
   \label{fig:limitLepPrime}
\end{figure}

Figure~\ref{fig:limitLepPrime} shows the exclusion limits for pair-produced $\tau'$ with 1e charge (left) and with 2e charge (right). 

\begin{figure}[ht!]
   \centering
       \includegraphics[width=.42\textwidth]{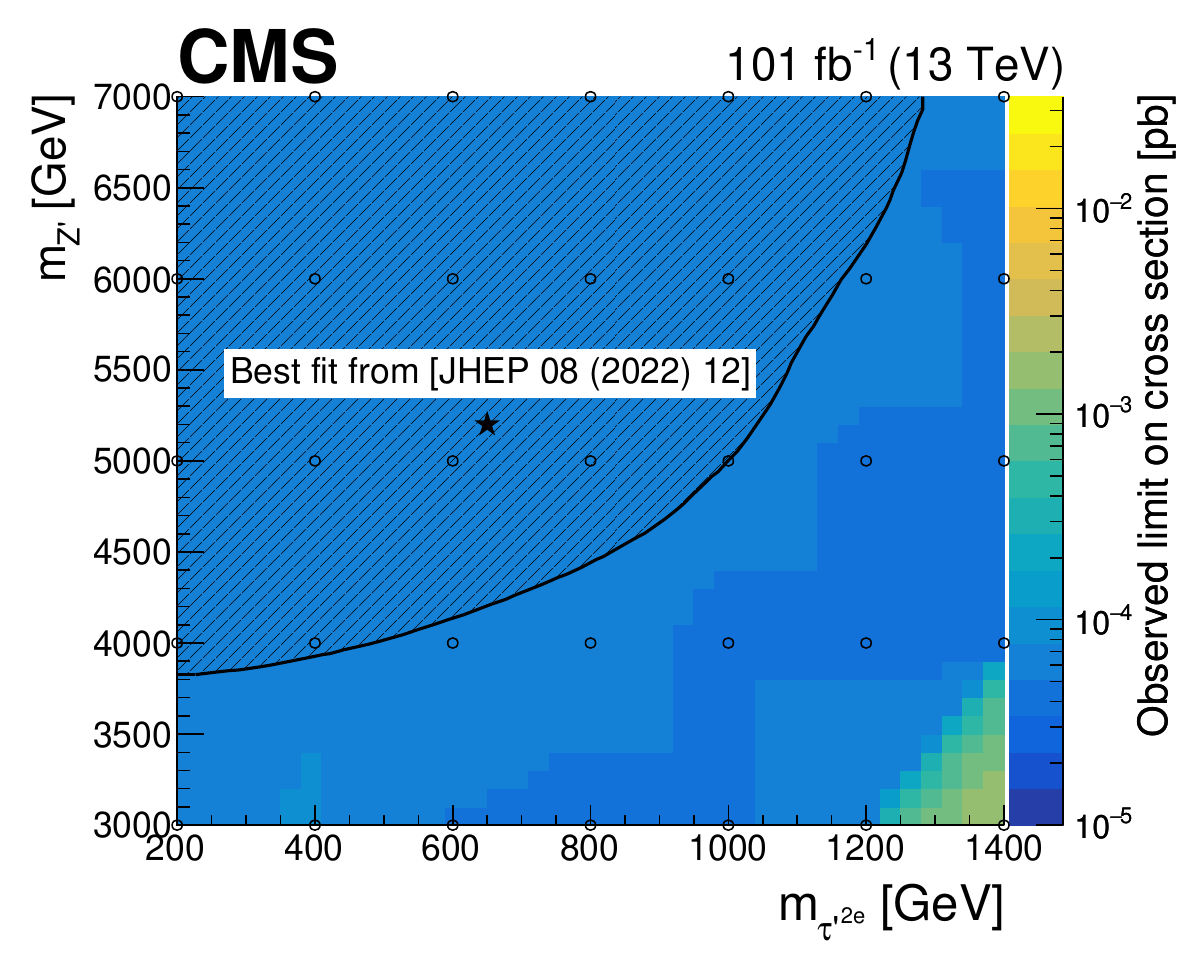}
       \includegraphics[width=.42\textwidth]{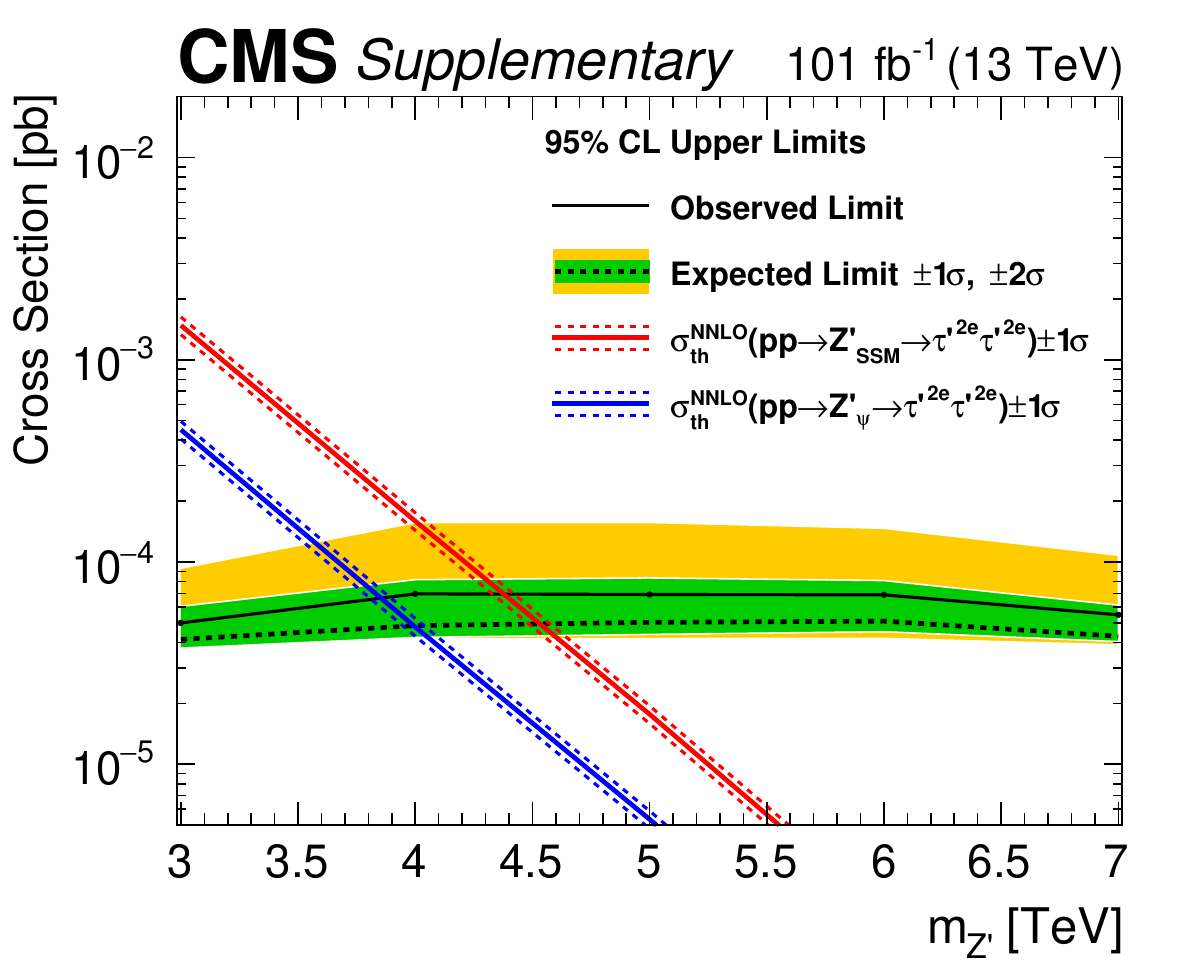}
        \caption{Exclusion limits for pair-produced stau (left) and GMSB produced stau (right).}
   \label{fig:limitZPrime}
\end{figure}

Figure~\ref{fig:limitZPrime} shows the exclusion limits for Z$'$-produced $\tau'$ with 1e charge (left) and with 2e charge (right). 

The mass method~\cite{Khachatryan:2016sfv} inverts Eq.~\ref{eq:MassFromHarmonicEstimator} to calculate the mass~\cite{EXO-12-026}  and performs a counting experiment in a mass windows for each interpretation and mass point. The full description of the method can be found in Ref.~\cite{cmscollaboration2024searchheavylonglivedcharged}. Table~\ref{tab:allResults} is the executive summary showing the expected and observed mass limits for various HSCP interpretations including both background estimate methods. The two methods show similar sensitivity. The single excess event in the ionization method makes the observed limits weaker, while for the case of the mass method there was no excess so the observed limit is in line with the  expected one.

\begin{table}[ht!]
  
   \label{tab:MassLimits} 
  \centering
  \begin{tabular}{|l|cc|cc|} 
    \hline
    Model       & \multicolumn{2}{c}{Ionization method} & \multicolumn{2}{|c|}{Mass method} \\
    & Exp. ({TeV}) & Obs. ({TeV}) & Exp. ({TeV})  & Obs. ({TeV}) \\ 
    \hline
    {gluino}  & 2.06 $\pm$ 0.06 & 2.06  & 2.08 $\pm$ 0.02  & 2.08  \\
    \hline
    {stop}  & 1.43 $\pm$ 0.05 & 1.40  & 1.47 $\pm$ 0.02  & 1.47  \\
    \hline
    {GMSB SPS7 stau} & 0.86 $\pm$ 0.07 & 0.85  & 0.87 $\pm$ 0.05  & 0.85  \\
    \hline
    pair-prod. stau-R & 0.53 $\pm$ 0.03 & 0.52  & 0.50 $\pm$ 0.07  & 0.51  \\
    \hline
    {pair-prod. stau-L}  & 0.66 $\pm$ 0.04 & 0.64  & 0.67 $\pm$ 0.06  & 0.61  \\
    \hline
    {pair-prod. stau-L/R} & 0.71 $\pm$ 0.04 & 0.69  & 0.75 $\pm$ 0.08  & 0.64  \\
    \hline
    {DY-prod $\tau'$ ($Q=1e$)}  & 1.05 $\pm$ 0.05 & 1.02  & 1.14 $\pm$ 0.03  & 1.14  \\
    \hline
    {DY-prod $\tau'$ ($Q=2e$)}  & 1.35 $\pm$ 0.05 & 1.32  & 1.41 $\pm$ 0.02  & 1.41  \\
    \hline
    {$Z$'$_{Phi} \to \tau' \tau'$ }  & 3.99 $\pm$ 0.21 & 3.95  & 4.03 $\pm$ 0.01 & 4.03  \\
    \hline
    {$Z$'$_{SSM} \to \tau' \tau'$ } & 4.53 $\pm$ 0.23 & 4.38  & 4.56 $\pm$ 0.01 & 4.57  \\ 
    \hline
  \end{tabular}
  \caption{Expected and observed mass limits for various HSCP interpretations including both background estimate methods.}
  \label{tab:allResults}
\end{table}

\section{Conclusions}

A signature based, model independent search for HSCPs has been presented. The background was studied with MC samples, and the preselection was optimized to remove both physically motivated highly ionizing SM processes together with mismeasured tracks.
Two methods of data-driven background predictions were developed. A novel approach relying on the independence of the ionization in the tracking detectors and an improved version of the historical mass method. This proceedings concentrates on the ionization method and includes the mass results for completeness. 
No significant excess over the SM expectation was found.
The results were interpreted in 10 different models,
one of them is a direct response to an excess seen by ATLAS.
With this the mass exclusion limits are best published to date.


\bibliography{main}
\bibliographystyle{elsarticle-num} 

\end{document}